\documentclass[12pt]{iopart}

\usepackage{iopams}
\usepackage{graphicx}
\usepackage{caption}

\newcommand{\dd}{\mathbf{d}}

\newcommand{\tensor}[2]{\mathbf{#1}_\mathrm{#2}}
\newcommand{\connection}[2]{\omega^{#1}_{\phantom{#1}#2}}
\newcommand{\curvature}[2]{\Omega^{#1}_{\phantom{#1}#2}}
\newcommand{\z}{z^\pm}
\newcommand{\ro}{\bar{\rho}}
\newcommand{\p}{\bar{p}_c}
\newcommand{\0}[1]{\mathcal{O}(\beta^{#1})}
\newcommand{\R}[1]{\mathcal{R}_{\mathrm{#1}}}
\newcommand{\Ss}{\mathcal{S}^2}

\begin{document}

\title[An exact static two-mass solution]{An exact static two-mass solution 
using Nariai spacetime}
\author{Michael Fennen$^1$ and Domenico Giulini$^{1,2}$}
\address{$^1$Center for Applied Space Technology and Microgravity (ZARM)\\ 
         University of Bremen, Germany}
\address{$^2$Institute for Theoretical Physics\\ 
         Riemann Center for Geometry and Physics\\ 
Leibniz University of Hannover, Germany}
\ead{michael.fennen@zarm.uni-bremen.de\\
giulini@itp.uni-hannover.de}

\begin{abstract}
We show the existence of static, spherically symmetric spacetimes 
containing two stars of incompressible matter, possibly oppositely 
charged. The stars are held apart by the negative pressure of a 
positive cosmological constant but there is no cosmological horizon 
separating them. The spacetime between the stars is given by the 
Nariai solution, or a slight generalisation thereof in the charged
case. 
\end{abstract}
%\pacs{}
%\submitto{\JPG}

\maketitle

% Intro
%%%%%%%%%%%%%%%%%%%%%%%%%%%%%%%%%%%%%%%%%%%%%%%%%%%%%%%%%%%%
\section{Introduction}
\label{sec:Itroduction}
Solutions to the Einstein equations with conventional matter 
(obeying reasonable energy conditions) representing two stars cannot be 
expected to be both, static and regular. After all, the stars 
will attract each other, so that without any agent 
keeping them apart they must inevitably start to approach. 
Staticity can only be enforced if one allows for singularities 
in the geometry outside the stars, usually either along the 
line connecting the stars and/or lines connecting each star to 
infinity (along the axis, in an axisymmetric situation). 
These singular lines can be interpreted as ``struts'' or 
``cords'' that by their compressive or tensile stresses 
hold the stars in place. Many solutions showing this structure 
are known; see \cite{Bach.Weyl:1922} and \S\,35 of \cite{Weyl:RZM1991}
for early discussions and chapter\,10 of \cite{Griffiths.Podolsky:ExactSpaceTimes} for a comprehensive modern text-book account. 

Alternatively, one may ask whether the stars could be held
apart by the large negative pressure of a positive cosmological 
constant. Clearly, such a scenario is not meant to apply to 
realistic pairs of stars in our universe, but as a matter of 
principle concerning the study of exact solutions to Einstein's 
equation is seems an obvious question to ask. 

In fact, already in 1922 the mathematician and engineer Erich 
Trefftz attempted to find a static solution to Einstein's 
vacuum equations with cosmological constant representing two 
``point masses''\cite{Trefftz:1922}. That solution was, in fact, 
locally identical to the Schwarzschild - De\,Sitter solution, 
also known as Kottler solution (due to \cite{Kottler:1918}), 
that we now interpret as a single, spherically symmetric 
uncharged black hole in the De\,Sitter universe. Trefftz' 
interpretation was 
different. He looked upon this solution as describing two 
``point masses'' (i.e. black holes) placed at antipodal 
points of a 3-sphere. In this way he could maintain full 
spherical symmetry for the two-body situation, and not 
just axisymmetry, as would be the case if the two masses 
where not placed at antipodal points of the Universe. Einstein 
immediately reacted to Trefftz' paper \cite{Einstein:CP13} 
(Doc.\,387, pp.\,595-596) by pointing out that the surface area 
of the spheres of symmetry (i.e. the $SO(3)$ orbits) must 
assume a stationary value somewhere in the region exterior 
to the stars and that this implies, according to the field 
equations, that the time-time component of the metric assumes 
the value zero. This, according to Einstein's interpretation,
was the signal of a true and intolerable singularity.%
\footnote{Quite surprisingly, initially even Hermann Weyl 
in \cite{Weyl:1919} followed Einstein's belief that a non-vanishing cosmological constant would prevent the existence of vacuum 
solutions (``es widerstreitet dem 
Einstein'schen Gesetz, dass die Welt vollst\"andig masseleer ist''; \cite{Weyl:1919}) and that in static spacetimes with Killing vector 
$\partial/\partial t$ the vanishing of 
$\tensor{g}{}(\partial/\partial t,\partial/\partial t)$ signals 
a singularity.}

Today we know better: Einstein's ``singularity'' is a mere
Killing horizon that shows the existence of a cosmological 
event horizon (in agreement with Hawking's strong rigidity theorem)
in the region between the two bodies. In the present paper we 
wish to address the question of whether we can use the
cosmological constant to support to massive stars at finite 
distance in static equilibrium \emph{without} any cosmological   
horizon separating them.

\section{Spherical symmetry}
In order to evade the conclusions of Einstein's argument we 
need to identify its mathematical origin. This is easy to do 
if one takes due care 
of the restrictions imposed by spherical symmetry. Recall 
that a spacetime $(M,g)$ is spherically symmetric if there 
exists an action of $SO(3)$ by isometries whose orbits are 
spacelike 2-spheres. These spheres can then be parametrised 
in the usual fashion by spherical polar coordinates $\vartheta$
and $\varphi$. Moreover, the normal bundle to these orbits 
can be shown to be integrable (a step usually omitted in 
most textbooks; an exception is  \cite{Straumann:GR2013}, 
section\,4.10.1). This means that the distribution of 
(Lorentzian) 2-planes perpendicular to the distribution of 
(Euclidean) 2-planes tangent to the $SO(3)$ orbits are 
locally integrable. Consequently, the metric can be 
locally parametrised  by coordinates $(t,r,\vartheta,\varphi)$
with no off-diagonal metric coefficients between $(t,r)$
and $(\vartheta,\varphi)$. 

Note that spherical symmetry alone implies other useful 
structural elements. For example, there is a preferred 
function, $R$, whose value at each point $p\in M$ 
equals $\sqrt{A(p)/4\pi}$, where $A(p)$ is the area 
of the $SO(3)$ orbit through $p$. This function $R$ is
called the \emph{areal radius}.  We can use $R$ as radius 
coordinate in regions where the one-form $\dd R$ nowhere 
vanishes. Suppose this being the case and that $\dd R$ is 
spacelike, i.e. the 3-dimensional sub-manifolds of constant 
$R$ are timelike. Then, up to normalisation, there is a unique 
$SO(3)$-invariant timelike vector field (necessarily orthogonal
to the $SO(3)$ orbits) that is annihilated by $\dd R$. It is 
not difficult to see that this vector field is hypersurface 
orthogonal and hence gives rise to a preferred foliation of 
spacetime by spacelike hypersurfaces.%
\footnote{This vector field is sometimes referred to as 
\emph{Kodama field}; e.g.,~\cite{Carrera.Giulini:2010a}.} 
Choosing a time function $t$ whose level sets are these hypersurfaces, 
the metric takes the form 
\begin{equation}
\label{eq:SpehricalSymmMetric-1}
\tensor{g}{}\, =-a(t,R)\,\dd t^2+r(t,R)\,\dd R^2+R^2(\dd\,\vartheta^2+\sin^2\vartheta\,\dd\varphi^2)\,.
\end{equation}
Similarly, the other two cases for non-vanishing $\dd R\ne 0$, 
where $\dd R$ is time- or lightlike, can also be written down.

Often \eref{eq:SpehricalSymmMetric-1} (together with the other 
two cases) are taken to exhaust the ``general forms'' of 
spherically symmetric metrics. But there is still the case 
left where $\dd R$ is neither spacelike, timelike, or 
lightlike, but simply vanishes, at least locally. In this 
case we cannot use $R$ as coordinate. 
Let us now focus on these somewhat exceptional cases and 
see under what conditions they occur as solutions to Einstein's 
equations, restricting attention to the static case. Then the 
metric can then be written in the form 
\begin{equation}
\label{eq:SpehricalSymmMetric-2}
\tensor{g}{}=- a^2(z)\, \dd t^2
+\dd z^2+R^2(z)\bigl(\dd \vartheta^2+\sin^2\vartheta\, \dd\varphi^2\bigr)\,,  
\end{equation}
where we now used a radial coordinate $z$ (called $z$ rather than 
$r$ for reasons to become clear soon) such that $\tensor{g}{}(\partial_z,\partial_z)=1$. The areal radius $R$ is 
now a function of $z$ that may well have stationary points. 
The non-vanishing components of the left-hand side of 
Einstein's equations, $G_{\mu\nu}+\Lambda g_{\mu\nu}$, with 
respect to the orthonormal co-frame 

\numparts
\label{eq:OrthonormalFrame}
\begin{eqnarray}
\label{eq:OrthonormalFrame-0}
\theta^0&=&a\dd t\,,\\ 
\label{eq:OrthonormalFrame-1}
\theta^1&=&\dd z\,,\\
\label{eq:OrthonormalFrame-2}
\theta^2&=&R\,\dd\vartheta\,,\\
\label{eq:OrthonormalFrame-3}
\theta^3&=&R\, \sin\vartheta\dd\varphi\,,
\end{eqnarray}
\endnumparts
are then given by (see Appendix \eref{eq:EinsteinTensorComp-00}-\eref{eq:EinsteinTensorComp-22}
and note that $1=-g_{00}=g_{11}=g_{22}=g_{33}$)
\numparts
\label{eq:Nariai-EinstEq}
\begin{eqnarray}
\label{eq:Nariai-EinstEq-00}
G_{00}-\Lambda&\ =&
-2\frac{R''}{R}+\frac{1-{R'}^2}{R^2}\,-\,\Lambda\,,\\
\label{eq:Nariai-EinstEq-11}
G_{11}+\Lambda&\ = &\!\!\quad
2\,\frac{a'R'}{aR}-\frac{1-{R'}^2}{R^2}\,+\,\Lambda\,,\\
\label{eq:Nariai-EinstEq-22}
G_{22}+\Lambda&\ = &\quad
\frac{a''}{a}+\frac{R''}{R}+\frac{a'R'}{aR}\,+\,\Lambda\,.
\end{eqnarray}
\endnumparts
We did not write down the $33$-component, which is identical 
to the $22$-component due to spherical symmetry.  

Let us focus on solutions to Einstein's equations with 
cosmological constant and vanishing energy-momentum tensor 
(vacuum solutions). Einstein's equations are then equivalent 
to each of the three expressions above being equal to zero.
Taking the sum of the first two expressions 
\eref{eq:Nariai-EinstEq-00} and  \eref{eq:Nariai-EinstEq-11}
and equating it to zero gives 
\begin{equation}
\label{eq:Nariai-EinstEq-Lemma}
aR''=a'R'\,.
\end{equation}
Now suppose $z=z_*$ is a stationary point for $R$, 
i.e. $R'(z_*)=0$. Then \eref{eq:Nariai-EinstEq-Lemma} shows 
that $a(z_*)=0$ if $R''(z_*)\ne 0$, i.e. if $R$ assumes a 
proper extremal value at $z_*$. But zeros of $a$ correspond 
to Killing horizons, which is just Einstein's observation 
(in modern terminology and interpretation). But now it is 
also clear how to avoid this conclusion (of a Killing horizon), 
namely to assume that $R$ is constant; $R=\R{N}$.
Either of \eref{eq:Nariai-EinstEq-00} or 
\eref{eq:Nariai-EinstEq-11} then gives
\begin{equation}
\label{eq:Nariai-Radius}
\R{N}=1/\sqrt{\Lambda}\,,
\end{equation}
which shows that we have to assume $\Lambda>0$.
The final equation \eref{eq:Nariai-EinstEq-22} gives 
$a''=-\Lambda a$, which is the harmonic-oscillator 
equation. The two integration constants (amplitude and phase) 
can be absorbed by redefining the scale of the $t$ and 
the origin of the $r$ coordinate. This leads to the 
\emph{Nariai} metric (in static form), known since 
1950 \cite{Nariai:1999-a}\cite{Nariai:1999-b}:
\begin{equation}
\label{eq:Nariai-Metric}
\tensor{g}{}=-\cos^2\bigl(z/\R{N}\bigr)\, \dd t^2+\dd z^2+
\R{N}^2(\dd \vartheta^2+\sin^2\vartheta\,\dd \varphi^2)\,.
\end{equation}
Note that the $SO(3)$ orbits are mutually isometric 2-spheres 
of radius $\R{N}$. This is why we called the spatial coordinate 
perpendicular to these orbits $z$ rather than $r$, because the
$z$-family of 2-spheres looks like a cylinder 
$\mathbb{R}\times S^2$. If we wish to avoid Killing horizons 
we have to restrict the cylinder to that region parametrised by 
$z\in(-\R{N}\pi/2,\R{N}\pi/2)$. 

Finally, if we consider the Einstein's equations with matter, 
\begin{equation}
\label{eq:Einstein-Eq}
G_{\mu\nu}+g_{\mu\nu}\Lambda=8\pi G T_{\mu\nu}\,,
\end{equation}
the sum of the $00$ and $11$ components \eref{eq:Nariai-EinstEq-00} 
and  \eref{eq:Nariai-EinstEq-11} tell us that at a stationary 
point $z_*$ for $R$ we have $8\pi G(T_{00}+T_{11})=-2R''/R$.
The weak-energy condition implies that the left-hand side is 
non-negative, hence $R''(z_*)\leq 0$. This implies that $R$ 
cannot have a local minimum inside a star whose matter obeys 
the weak energy condition. 

%Charged Nariai
%%%%%%%%%%%%%%%%%%%%%%%%%%%%%%%%%%%%%%%%%%%%%%%%%%%%%%%%%%%%

\section{Charged-Nariai spacetime} 
\label{sec:ChargedNariai}
In this section we will briefly show how to generalise the 
Nariai spacetime (\ref{eq:Nariai-Metric}) so as to include 
an electric field parallel to the cylinder axis which 
makes the two ends of the cylinder appear equally and 
oppositely charged. Note that we do not just solve Maxwell's 
equations on the background~(\ref{eq:Nariai-Metric}), but 
we seek a new solution to the Einstein-Maxwell equations 
that, in an appropriate sense, generalises 
(\ref{eq:Nariai-Metric}). That generalised metric shall be 
of the form (cf.~(\ref{eq:OrthonormalFrame}))
\begin{eqnarray} 
\tensor{g}{N} 
&=& -\theta^0\otimes\theta^0
 +\sum_{a=1}^3\theta^a\otimes\theta^a\nonumber \\
\label{eqn:NariaiAnsatz}
&=& - a^2(z) \,\dd t^2 + \dd z^2 + \R{N}^2 \,\dd \Omega^2\,,
\end{eqnarray}
with a yet unspecified function $a(z)$. The electric field is also 
assumed to be spherically symmetric and spacelike , which implies 
that tangent to the simultaneity hypersurfaces $dt=0$ it points 
parallel to the normal of the 2-sphere orbits of $SO(3)$ and that 
its modulus depends only on $z$. Hence, without loss of 
generality, the electromagnetic 2-form is given by 
\begin{equation} 
\label{eqn:NariaiEmTensor}
 \tensor{F}{} = - E(z) \,a(z) \,\dd t \wedge \dd z = \frac{q}{d^2(z)} \,a(z) 
 \,\dd t \wedge \,\dd z.
\end{equation}
Here $q$ is a constant and $d$ is a real-valued function with 
physical dimension of length. The signs are chosen such that 
positive $q$ correspond to electric fields pointing in negative
$z$ direction, as will become apparent below.   

In the absence of sources Maxwell's equations read 
$\dd \tensor{F}{} = 0$ and $\dd \ast \tensor{F}{} = 0$. 
The first equation is solved by $\tensor{F}{} = \dd \tensor{A}{}$
with $\tensor{A}{}=-\Phi(z)\,\dd t$ and 
\begin{equation}
 \Phi(z) = \int_{z_0}^z \frac{q}{d^2(x)} \,a(x) \,\mathrm{d} x\,.
\end{equation}
The Hodge-duality map $\ast$ is defined with respect to the 
space-time orientation represented by the volume form 
$\theta^0\wedge\theta^1\wedge\theta^2\wedge\theta^3$, so that, 
e.g., $\ast(\theta^0\wedge\theta^1)=-\theta^2\wedge\theta^3$. 
Hence 
\begin{equation}
 \ast \tensor{F}{} = -\frac{q}{d^2(z)} \,\R{N}^2 \sin \vartheta \,\dd 
 \vartheta 
 \wedge \dd \varphi = \frac{q}{d^2(z)} \,\R{N}^2 \,\dd(\cos \vartheta \,\dd 
 \varphi)\,.
\end{equation}
This shows that Maxwell's second equation is equivalent to $d(z)$ 
being constant. Since the two constants $q$ and $d$ only appear 
in the combination $q/d^2$, we may without loss of generality 
set $d(z) \equiv \R{N}$. The flux through any of the $SO(3)$ orbits 
therefore equals 
\begin{equation}
 Q = \frac{1}{4 \pi} \int_{S^2} \ast \tensor{F}{} = \pm q\,,
\end{equation}
where the sign on the right-hand side depends on the 
orientation of $S^2$. If the orientation of spacetime is 
represented by the volume form $\theta^0\wedge\theta^1\wedge\theta^2\wedge\theta^3$,  
the orientation of space (the level sets of  $t$) is represented 
by $\theta^1\wedge\theta^2\wedge\theta^3$ if $\partial/\partial t$
is taken to point \emph{outward} in spacetime. Furthermore, the 
orientation of the $SO(3)$ orbits is represented by 
$\theta^2\wedge\theta^3$ if $\partial/\partial z$ is taken to 
point \emph{outward} in space. Then $Q=-q$ which for $q>0$ 
means that the negative end ($z<0$) is negatively and the 
positive end ($z>0$) is positively charged. Accordingly, 
the electric field points into the negative $z$ direction, as
stated above.

\subsection{Einsteins's equations}
Next we evaluate Einstein's equation, using the ansatz 
\eref{eqn:NariaiAnsatz} for its left-hand side and 
\eref{eqn:NariaiEmTensor} for its right-hand side. 
The right-hand side  is given by the stress-energy tensor 
of an electromagnetic field, which is 
\begin{equation}
 T^{\mathrm{(em)}}_{\mu\nu} =\frac{1}{4\pi} \left( g^{\rho\sigma} F_{\mu\rho} 
 F_{\nu\sigma} - \frac{1}{4} g_{\mu\nu} F_{\rho\sigma} F^{\rho\sigma} \right)\,,
\end{equation}
where $F_{\mu\nu}$ are the components of the electromagnetic 
field tensor \eref{eqn:NariaiEmTensor}. With respect to the 
frame (\ref{eq:OrthonormalFrame}) the only non-vanishing 
component is $F_{01}=q/\R{N}^2$. Hence the electromagnetic 
stress-energy tensor has no off-diagonal components and reads  
$(q/\R{N}^2)^2(8\pi)^{-1}\,\mathrm{diag}(1,-1,1,1)$. Writing 
$8\pi G$ times these on the right-hand side of 
\eref{eq:Nariai-EinstEq-00} and \eref{eq:Nariai-EinstEq-22}
(due to $R'=R''=0$ the $00$- and $11$ components become identical)
Einstein's equations turn out to be identical to the following two 
equations:
\numparts
\begin{eqnarray}
 \frac{1}{\R{N}^2} - \Lambda = \frac{G q^2}{\R{N}^4}, 
 \label{eqn:NariaiEinsteinA} \\
 \frac{a''(z)}{a(z)} + \Lambda = \frac{G q^2}{\R{N}^4}. 
 \label{eqn:NariaiEinsteinB}
\end{eqnarray}
\endnumparts
The first equation \eref{eqn:NariaiEinsteinA} merely relates 
the three constants $\Lambda$, $\R{N}$, and $q^2$. In particular, 
it allows to express the cosmological constant $\Lambda$ in 
terms of the radius $\R{N}$ of the orbit 2-spheres and the 
charge $q$. Using this to eliminate $\Lambda$ in 
\eref{eqn:NariaiEinsteinB} we get 
\begin{equation} 
\label{eqn:NariaiDiffEquation}
 a''(z)+\frac{1}{\R{N}^2}
\left( 
1 - \frac{2G q^2}{\R{N}^2} 
\right) 
a(z)=0\,.
\end{equation}
Depending on the ratio of the charge to the radius, the solutions 
to \eref{eqn:NariaiDiffEquation} are: 
\begin{equation} \label{eqn:NariaiFunction}
 a(z) = \cases{
 \mathcal{C} \cos \left( \frac{\lambda z}{\R{N}} \right) + \mathcal{D} \sin 
 \left( \frac{\lambda z}{\R{N}} \right) 
 &if $q^2 < \frac{\R{N}^2}{2G}$\,, \\
 \mathcal{C} + \mathcal{D} z
 &if $q^2 = \frac{\R{N}^2}{2G}$\,, \\
 \mathcal{C} \cosh \left( \frac{\lambda z}{\R{N}} \right) + \mathcal{D} 
 \sinh 
 \left( \frac{\lambda z}{\R{N}} \right) 
 &if $q^2 > \frac{\R{N}^2}{2G}$\,,\\
 }        
\end{equation}
where $\lambda^2 = \left| 1 - \frac{2Gq^2}{\R{N}^2} \right|$ 
and $\mathcal{C}$ and $\mathcal{D}$ are integration constants. 

In the following we will choose $\mathcal{D} = 0$ so as to make 
the solutions invariant under the reflection $z \mapsto -z$. 
Without loss of generality we may then further set 
$\mathcal{C}=1$, since this can always be achieved by a 
constant rescaling of the time coordinate. The metric 
\begin{equation}
 \tensor{g}{N} = - \cos^2 \left( \frac{\lambda z}{\R{N}} \right) \dd t^2 + 
 \dd 
 z^2 + \R{N}^2 \,\dd \Omega^2
\end{equation}
may the be called a charged-Nariai spacetime. It reduces to 
the ordinary Nariai spacetime in the limit of vanishing 
charge and will play an important r{\^o}le  in what is to 
follow. Killing horizons are now absent if we restrict to 
$z\in(-\R{N}\pi/2\lambda,\R{N}\pi/2\lambda)$. 

The third case,
\begin{equation}
 \tensor{g}{BR} = - \cosh^2 \left( \frac{\lambda z}{\R{N}} \right) \dd t^2 + 
 \dd  z^2 + \R{N}^2 \,\dd \Omega^2\,,
\end{equation}
is the Bertotti-Robinson metric \cite{Bertotti:1959} generalised to 
non-zero cosmological constant, sometimes also referred to as 
cosmological Bertotti-Robinson spacetime~\cite{Griffiths.Podolsky:ExactSpaceTimes}. 
The special value $q^2=\frac{\R{N}^2}{G}$, i.e. $\lambda=1$, corresponds to vanishing cosmological constant.

Finally we remark on the notion of ``mass''. In spherically 
symmetric spacetimes there is, next to the areal radius $R$, 
another geometrically defined function, which is the sectional 
curvature of spacetime tangent to the $SO(3)$ orbits. (Note 
that this is generally \emph{not} the Gaussian curvature of 
the orbit, unless the orbit is totally geodesic.) Since the 
orbits foliate spacetime, this defines a function on spacetime 
by assigning to each point the sectional curvature tangent 
to the orbit through it. Clearly this function is constant 
on each orbit. Writing the metric in the form 
\eref{eq:SpehricalSymmMetric-2} this function is just given 
by the $R_{23\,23}$ component of the Riemann tensor with respect 
to the frame \eref{eq:OrthonormalFrame-0}-\eref{eq:OrthonormalFrame-3},
which we calculated in~\eref{eq:RiemanntensorComp-23}. Note that 
the sectional curvature of spacetime tangent to the group orbits 
equals the Gaussian curvature of the 2-dimensional surface locally 
spanned by all spacetime geodesics starting tangentially to the 
orbit. This surface touches the orbit to first, but generally not to 
second order. Consequently, their Gaussian curvatures are generally 
not the same, as can be clearly seen from 
expression~\eref{eq:RiemanntensorComp-23}, which differs from $R^{-2}$
(the Gaussian curvature of the orbits) by the $-R'^2/R^2$ term,
which is non zero iff $R'\ne 0$. Hence, in the case of constant $R$,
the spacetime's  sectional curvature tangent to the orbits is 
identical to their Gaussian curvature.  

Now, if we multiply this sectional curvature by half the third power 
of the areal radius we get another function on spacetime which 
is constant on the orbits and which equals the Misner-Sharp mass
in geometric units. (The Misner-Sharp mass was first introduced 
in a non-geometric fashion in \cite{Misner.Sharp:1964}. 
Its geometric definition is discussed, e.g., in 
\cite{Carrera.Giulini:2010a}). It can be shown 
\cite{Carrera.Giulini:2010a} to equal the Hawking 
mass~\cite{Hawking:1968} and, according to  
\eref{eq:RiemanntensorComp-23}, has the following simple form 
(in physical units where $c=1$)
\begin{equation}
\label{eq:MisnerSharpMass}
 M =\frac{R}{2G}\Bigl(1-g^{-1}(\dd R,\dd R)\Bigr)=\frac{\R{N}}{2G}.
\end{equation}
where the first expression is the generally valid one 
if $R$ denotes the areal radius, and the second expression is 
valid for $dR=0$. We conclude that the charged-Nariai spacetime 
is almost determined by the mass and charge. ``Almost'' but not quite completely, because we also need to 
indicate the range of the coordinate $z$, i.e. the length of the 
cylinder. This makes the charged Nariai solution a three-parameter 
family. 
  
A star matched to the Nariai metric must have the very same 
mass as a result of the matching conditions. Two stars matched to 
the Nariai cylinder, one at each end, must clearly have equal 
and opposite charge. Once the mass and charge are fixed the 
only degree of freedom that is left is the position of the 
stars, i.e., the length of the cylinder. We expect that the 
position should be related to the pressure within the star, 
a property not yet used. To see this, we first need to find 
the star's interior solutions.

%Charged Star
%%%%%%%%%%%%%%%%%%%%%%%%%%%%%%%%%%%%%%%%%%%%%%%%%%%%%%%%%%%%
\section{Charged star solution} \label{sec:ChargedStar}

In this section we wish to derive simple solutions for spacetime 
regions interior to the star. We will take the distribution of
bare rest-mass, $\rho$, to be constant (with respect to the proper 
geometric measure induced in the hypersurfaces of simultaneity) 
and the charge distribution similarly simple, though not proportional 
to the distribution of bare rest-mass. So our solution will be a 
generalisation of the inner Schwarzschild solution 
\cite{Schwarzschild:1916b} to the charged case including a 
cosmological constant. Both cases have already been considered 
separately before: the neutral case with a cosmological constant 
in \cite{Boehmer:2003,Boehmer:2004,BoehmerFodor:2008}, 
and the charged case without cosmological constant in 
\cite{Kyle.Martin:1967}. To our knowledge, the only treatment 
of spherically-symmetric static stars with charge and non-vanishing 
cosmological constant is given in \cite{Boehmer:2011}. 
But the equations of state used in this paper differ from our 
condition of constant mass-density (incompressibility). Our intended application of this solutions is also different.  

As we have already seen, $R$ cannot assume a local minimum 
inside a star made of matter satisfying the weak energy-condition. 
As the centre of the star is a fixed point of the action of 
$SO(3)$, and the metric is required to be regular inside the star,
$R$ must tend to zero as we approach the centre. Hence $R$ is a 
monotonic function as the radius increases from zero until the 
first maximum is reached.%
\footnote{In Appendix\,B of \cite{Hawking.Ellis:TLSSOS} is 
allegedly shown that $R$ cannot have any extremum inside 
spherically-symmetric and static stars satisfying the weak 
energy-condition, but that is not correct. It seems that 
the existence of a term corresponding to our $R''(z_*)$
($Y''$ in their notation, resulting from their equation (A3)) 
has been overlooked.}
As for our construction we will only be interested in 
stars where $R$ assumes a maximum on its boundary, we may 
use $R$ as a coordinate function inside the star, which 
we now call $r$. Hence the metric inside the star may be 
written in the form  
\begin{eqnarray} 
\label{eqn:StarAnsatz}
\tensor{g}{S} 
&=&-\theta^0\otimes\theta^0+\sum_{a=1}^3\theta^a\otimes\theta^a\nonumber \\
&=&=-\e^{2a(r)} \,\dd t^2 + \e^{2b(r)} \,\dd r^2 + r^2 \,\dd \Omega^2\,,
\end{eqnarray}
where now 
\numparts
\label{eq:OrthonormalFrameStar}
\begin{eqnarray}
\label{eq:OrthonormalFrameStar-0}
\theta^0&=&\e^{a(r)} \,\dd t\,,\\ 
\label{eq:OrthonormalFrameStar-1}
\theta^1&=&\e^{b(r)} \,\dd r\,,\\
\label{eq:OrthonormalFrameStar-2}
\theta^2&=&r\,\dd\vartheta\,,\\
\label{eq:OrthonormalFrameStar-3}
\theta^3&=&r\, \sin\vartheta\dd\varphi\,.
\end{eqnarray}
\endnumparts

\subsection{Maxwell equations}
The most general static and spherically-symmetric electric field 
is given by the electromagnetic 2-form
\begin{equation} 
\label{eqn:StarEmField}
 \tensor{F}{}= 
-E(r) \,\theta^0\wedge\theta^1=
-E(r) \,\e^{a(r)+b(r)} \,\dd t \wedge \dd r\,.
\end{equation}
As before, Maxwell's first equation, $\dd \tensor{F}{} = 0$, 
is solved by $\tensor{F}{}=\dd \tensor{A}{}$, where 
$\tensor{A}{} = - \Phi(r) \,\dd t$ with 
\begin{equation}
 \Phi(r) = -\int_0^r \mathrm{d} x \,E(x) \,\e^{a(x)+b(x)} \,.
\end{equation}
Maxwell's second (inhomogeneous) equation reads  
$\dd \ast \tensor{F}{} = 4\pi \ast \tensor{J}{}$, where  
$\tensor{J}{} = -\sigma(r)\,\theta^0$ is the current-density 
1-form. On the left-hand side we get 
\begin{equation}
 \dd \ast \tensor{F}{} = -\frac{\mathrm{d}}{\mathrm{d} r} \left( E(r) \,r^2 
 \right) \dd r \wedge \dd \left( \cos \vartheta \,\dd \varphi \right)
\end{equation}
and on the right-hand side 
\begin{equation}
 4\pi\,\ast \tensor{J}{} = 
 4\pi\,\sigma(r)\,\theta^1\wedge\theta^2\wedge\theta^3 = 
- 4\pi\,\sigma(r) \,\e^{b(r)} \,r^2 \,\dd r \wedge \dd \left(  
 \cos \vartheta \,\dd \varphi \right).
\end{equation}
Hence the inhomogeneous Maxwell equations are equivalent to 
\begin{equation} \label{eqn:ElectricFieldDiff}
 \frac{\mathrm{d}}{\mathrm{d} r}\left( E(r) r^2 \right) = 4 \pi r^2 \sigma(r) 
 \,\e^{b(r)}\,,
\end{equation}
the solution of which is readily obtained if we restrict to a 
particular radial charge distribution given by 
\begin{equation}
\label{eq:ChargeDistribution} 
\sigma(r) \,\e^{b(r)} = \sigma_\pm = \mathrm{const.}
\end{equation}
Note that this does not correspond to constant charge 
density with respect to the proper (3-dimensional) geometric volume 
measure $\theta^1\wedge\theta^2\wedge\theta^3$, but rather to a 
constant density with respect to the ``areal volume'' measure 
$r^2\sin\vartheta\,\dd r\wedge\dd\vartheta\wedge\dd\varphi$. 
We will later see that, assuming a constant bare rest-mass 
distribution with respect to the proper geometric volume,
the (active) gravitational mass (which takes into account 
gravitational binding energies) will also be constantly distributed with 
respect to the ``areal volume'' (compare equation \eref{eq:MisnerSharpInsideStar-2}). Hence 
\eref{eq:ChargeDistribution} amounts to the assumption that 
the densities for gravitational mass and electric charge are 
constant inside the star. The charge inside a ball of areal 
radius $r$ now becomes 
\begin{equation} 
\label{eqn:ChargeFunction}
 Q(r) = 4\pi \int_0^r \mathrm{d} x \, x^2 \,\sigma(x) \,\e^{b(x)} = 
 \frac{4\pi}{3} \sigma_\pm r^3
\end{equation}
and the solution to \eref{eqn:ElectricFieldDiff} is then, clearly, 
just
\begin{equation} 
\label{eqn:ElectricField}
 E(r) = \frac{Q(r)}{r^2}.
\end{equation}

\subsection{Einstein's equations}

We recall that the components of the electromagnetic stress-energy 
tensor are given by 
\begin{equation}
T^{\mathrm{(em)}}_{\mu\nu} =\frac{1}{4\pi} \left( g^{\rho\sigma} F_{\mu\rho} 
F_{\nu\sigma} - \frac{1}{4} g_{\mu\nu} F_{\rho\sigma} F^{\rho\sigma} \right)\,,
\end{equation}
In what follows, all components refer to the orthonormal basis \eref{eq:OrthonormalFrameStar-0}-\eref{eq:OrthonormalFrameStar-3}.  
In the electric and spherically-symmetric case only the $F_{01}$
component is non-zero and given by \eref{eqn:ElectricField}, 
so that
\begin{equation}
\label{eq:EnergyMomentumComponents}
T_{\mu\nu}=\frac{Q^2(r)}{8\pi\,r^4}\mathrm{diag}(1,-1,1,1)\,. 
\end{equation}
For the metric \eref{eqn:StarAnsatz} and the energy-momentum 
Tensor above the $00$, $11$, and $22$ components of Einstein's 
equation contain all the information. They read, respectively, 
(dropping arguments of functions for brevity)
\numparts
\begin{eqnarray}
 \frac{1}{r^2} + \left( \frac{2b'}{r} -\frac{1}{r^2} \right) \e^{-2b} -\Lambda 
 = 2 \ro + \frac{G Q^2}{r^4}, \label{eqn:StarEinsteinA} \\
 -\frac{1}{r^2} +\left( \frac{2a'}{r} +\frac{1}{r^2} \right) \e^{-2b} +\Lambda 
 = 2 \bar{p} - \frac{G Q^2}{r^4}, \label{eqn:StarEinsteinB} \\
 \left( a'' + a'^2 - a'b' + \frac{a'-b'}{r} \right) \e^{-2b} +\Lambda 
 = 2 \bar{p} + \frac{G Q^2}{r^4}\,. \label{eqn:StarEinsteinC}
\end{eqnarray}
\endnumparts
Here and in the sequel we used the shorthand 
\begin{equation}
\label{eq:BarShorthand}
\bar{X}:= 4\pi G X\,,\quad (X=p,\rho,\sigma_\pm)\,.
\end{equation} 
Stress-energy conservation, $\nabla_\mu T^{\mu\nu}=0$, is 
equivalent to 
\begin{equation} \label{eqn:StarConservation}
 \bar{p}'(r) + (\ro + \bar{p}(r)) \,a'(r) = \frac{G Q(r) Q'(r)}{r^4}.
\end{equation}
We can rewrite the first equation \eref{eqn:StarEinsteinA} as
\begin{equation}
 \frac{\mathrm{d}}{\mathrm{d} r} \left( r e^{-2b} \right) 
 = 1 - 2 \ro r^2 - \frac{\bar{\sigma}_\pm^2}{9 G}  r^4 - \Lambda r^2
\end{equation}
with the solution
\begin{equation} 
\label{eqn:ChargedBFunction}
 \e^{-2b} = 1- \alpha r^2 - \beta r^4\,,
\end{equation}
where
\numparts
\begin{eqnarray}
\label{eq:DefParameterAlpha}
\alpha &=& \frac{1}{3} (2 \ro + \Lambda)\,,\\
\label{eq:DefParameterBeta}
\beta&=&\frac{\bar{\sigma}^2_\pm}{45 G}\,.
\end{eqnarray}
\endnumparts 

Eliminating $p(r)$ from \eref{eqn:StarEinsteinC} using  \eref{eqn:StarEinsteinB} we get a differential equation 
for $a(r)$. Using $GQ^2(r) = 5\beta r^6$ and \eref{eqn:ChargedBFunction} we obtain
\begin{equation}
 (1- \alpha r^2 - \beta r^4) \left( a'' + {a'}^2 \right) - \frac{a'}{r} (1+ 
 \beta r^4) = 11 \beta r^2.
\end{equation}
This somewhat complicated non-linear second order differential equation
is simplified by substituting 
\begin{equation} \label{eqn:ChargedDiffNu}
 a'(r) = \frac{r}{\sqrt{1- \alpha r^2 - \beta r^4}} \left( \sqrt{11 \beta} + 
 \frac{1}{f(r)} \right)  
\end{equation}
after which it becomes an ordinary linear first-order 
differential equation
\begin{equation} \label{eqn:ChargedDiffF}
 f'(r) = \frac{r}{\sqrt{1- \alpha r^2 - \beta r^4}}(2 \sqrt{11 \beta} f(r)+1)\,.
\end{equation}
This is easily integrated by separation and yields
\begin{equation} 
\label{eqn:ChargedFunction}
2 \sqrt{11 \beta} f(r) = \mathcal{B} \exp \left( \sqrt{11} \arcsin \left( 
\frac{\alpha + 2 \beta r^2}{\sqrt{\alpha^2 + 4 \beta}} \right) \right) - 1\,.
\end{equation}
Here $\mathcal{B}$ is an integration constant which depends 
on the central pressure $p_c = p(0)$, as we will see in the 
next section. We can combine the two equations \eref{eqn:ChargedDiffNu} and \eref{eqn:ChargedDiffF} to
\begin{equation}
 a'(r) = \frac{\sqrt{11 \beta} + \frac{1}{f(r)}}{2 \sqrt{11 \beta} f(r) + 1} 
 f'(r)\,,
\end{equation}
which is also easily integrated to 
\begin{equation}
 \e^{2a(r)} = \frac{\mathcal{A} f(r)^2}{2 \sqrt{11 \beta} f(r) + 1}\,,
\end{equation}
with another integration constant $\mathcal{A}$. This constant could be 
absorbed by a rescaling of the time coordinate $t$ but we have to keep it here 
because we have already used the rescaling freedom in the 
charged-Nariai spacetime. 

Altogether we get for a charged star the metric
\begin{equation}
 \tensor{g}{S} = - \frac{\mathcal{A} f^2(r)}{2 \sqrt{11 \beta} f(r)+1} \,\dd 
 t^2 + \frac{1}{1-\alpha r^2- \beta r^4} \,\dd r^2 + r^2 \,\dd \Omega^2.
\end{equation}
The pressure function is determined by the second equation 
\eref{eqn:StarEinsteinB} as
\begin{equation} \label{eqn:ChargedPressure}
  \bar{p}(r) = \sqrt{1- \alpha r^2- \beta r^4} \left( \sqrt{11 \beta} +   
  \frac{1}{f(r)} \right) + \frac{\Lambda - \alpha}{2} + 2 \beta r^2.
\end{equation}

The radial coordinate $r$ is only valid for $r<\R{S}$ 
where $\R{S}$ is the first positive root of $\e^{-2b(r)}$. 
This means $1-\alpha\R{S}^2-\beta\R{S}^4=0$ so that there 
is a coordinate singularity in the metric at $r=\R{S}$. 
However, in the following it will be necessary that 
the metric is regular here because we will see that this 
will turn out to be the radius of our stars. Therefore 
we rearrange
\begin{equation}
 \e^{-2b(r)} = 1- \alpha r^2- \beta r^4 
= \beta \left((r^2 + \Ss) (\R{S}^2-r^2)\right)\,,
\end{equation}
where $2 \beta \Ss = \alpha + \sqrt{\alpha^2 + 4 \beta}$ 
and $ 2 \beta \R{S}^2 = -\alpha + \sqrt{\alpha^2 + 4 \beta}>0$. 
Using a new radial coordinate $\chi$ such that 
$r = \R{S} \sin \chi$ allows us to eliminate the coordinate 
singularity in the metric at the equator $\chi=\frac{\pi}{2}$ 
or $r = \R{S}$, respectively. For $0 \le \chi \le 
\frac{\pi}{2}$ we simply have $F(\chi) = f(\R{S} \sin \chi)$ or
\begin{equation}
 2 \sqrt{11 \beta} F(\chi) = \mathcal{B} \exp \left( \sqrt{11} \arcsin \left( 
 \frac{\alpha + 2 \beta \R{S}^2 \sin^2 \chi}{\sqrt{\alpha^2 + 4 \beta}} 
 \right) \right) - 1.
\end{equation}
However, it is possible to extend the solution beyond the equator up to the 
second pole $\chi = \pi$. For $\frac{\pi}{2} \le \chi \le \pi$ we get
\begin{equation}
 2 \sqrt{11 \beta} F(\chi) = \mathcal{B} \exp \left( \sqrt{11} \left( \pi - 
 \arcsin \left( \frac{\alpha + 2 \beta \R{S}^2 \sin^2 \chi}{\sqrt{\alpha^2 + 
 4 
 \beta}} \right) \right) \right) - 1.
\end{equation}
The derivative of $F$ is given by
\begin{equation} 
\label{eqn:DerivationF}
 F'(\chi) 
= \frac{\R{S} \sin \chi}{\sqrt{\beta \left( \R{S}^2 \sin^2 \chi 
 + \Ss \right)}} \left( 2 \sqrt{11 \beta} F(\chi) + 1 \right)
\end{equation}
is regular for the whole interval $\left[ 0, \pi \right]$. 
So this new coordinate covers the whole (distorted) 3-sphere 
with the metric 
\begin{equation}
\label{eq:MetricChi}
 \fl \tensor{g}{S} = - \frac{\mathcal{A} F^2(\chi)}{2 \sqrt{11 \beta} 
 F(\chi) + 1} \,\dd t^2 + \frac{1}{\beta \left( \R{S}^2 \sin^2 \chi + \Ss 
 \right)} \,\dd\chi^2 + \R{S}^2 \sin^2 \chi \,\dd \Omega^2.
\end{equation}
Finally, the pressure becomes
\begin{equation} \label{eqn:StarPressure} 
 \fl \bar{P}(\chi) = \sqrt{\beta \left( \R{S}^2 \sin^2 \chi + \Ss 
 \right)} \left( \sqrt{11 \beta} + \frac{1}{F(\chi)} \right) \R{S} \cos \chi 
 + \frac{\Lambda - \alpha}{2} + 2 \beta \R{S}^2 \sin^2 \chi.
\end{equation}

\goodbreak

Let us end this section with a few observations and remarks:
\begin{enumerate}
\item
The function $F(\chi)$ is strictly monotonic, increasing for 
$\mathcal{B}>0$ and decreasing for $\mathcal{B}<0$.  
\item
Using the general expression for the Misner-Sharp mass 
\eref{eq:MisnerSharpMass} applied to the metric 
\eref{eq:MetricChi}, from which the areal radius immediately 
follows to be $R=\R{S}\sin\chi$, and also taking into account the 
definitions \eref{eq:DefParameterAlpha}-\eref{eq:DefParameterBeta},
we get the following expression for the Misner-Sharp mass inside a 
ball of latitude $\chi$:
\begin{equation}
\label{eq:MisnerSharpInsideStar-1}
M(\chi)=\frac{\R{S}^3\sin^3\chi}{2G} 
\left(\alpha+\beta\R{S}^2\sin^2\chi\right).
\end{equation}
The first term in this equation comprises the contributions 
from the matter and the cosmological constant. If expressed 
directly in terms of the parameters $\rho$ and $\Lambda$ and also 
in terms of the areal radius $R=r$ (recall that in the coordinates 
in which the metric is written as \eref{eqn:StarAnsatz} we have 
$R=r$) it reads 
\begin{equation}
\label{eq:MisnerSharpInsideStar-2}
M_{(\rho,\Lambda)}(r)=
\frac{4\pi}{3}r^3\left(\rho+\frac{\Lambda}{8\pi G}\right)\,.
\end{equation}
Note that $\frac{4\pi}{3}r^3$ is \emph{not} the spatial 
volume $V(r)$ of the ball bound by the sphere of areal 
radius $r$. The latter is bigger and the difference   
$\bigl(\frac{4\pi}{3}r^3-V(r)\bigr)\rho$ just accounts 
for the (negative) gravitational binding energy of the 
matter represented by $\rho$. This is well known from 
the ordinary inner Schwarzschild solution.  Note that the 
contribution from the cosmological constant is likewise 
diminished by this volume factor. In fact, the very same is 
also true for the electromagnetic part: Using 
\eref{eq:DefParameterBeta}, \eref{eq:BarShorthand}, and
\eref{eqn:ChargeFunction} we can easily see that the 
second term in \eref{eq:MisnerSharpInsideStar-1} equals 
\begin{equation}
\label{eq:MisnerSharpInsideStar-3}
M_{Q}(r)=\frac{1}{10}\cdot\frac{Q^2(r)}{r}
\end{equation}
which is precisely the flat-space result for the 
energy stored in the electric field inside a 
homogeneously charged ball of radius $r$.
\item
At this stage this manifold of solutions has more free 
parameters (four) than the charged Nariai solution (three). 
But additional dependencies will be imposed on the former by 
the junction conditions, as we will discuss  in the following \sref{sec:TwoMass}. 
\item
The uncharged solutions, which are clearly obtained 
by integrating all differential equations after setting 
$\beta=0$, are also obtained from our solutions in the 
limit $\beta \rightarrow 0$. We note that this would not 
be true if we had chosen the Ansatz of \cite{Kyle.Martin:1967}.
However, in taking the limit $\beta\rightarrow 0$ one must 
be careful with the boundary conditions which may also 
depend on $\beta$. We will demonstrate this for our 
special case of the Nariai spacetime 
in \sref{sec:NeutralLimit} in detail.
\end{enumerate}

%Two Mass
%%%%%%%%%%%%%%%%%%%%%%%%%%%%%%%%%%%%%%%%%%%%%%%%%%%%%%%%%%%%

\section{Two-mass solution} \label{sec:TwoMass}

\subsection{Junction conditions}

\begin{figure}[b]
 \centering
 \input{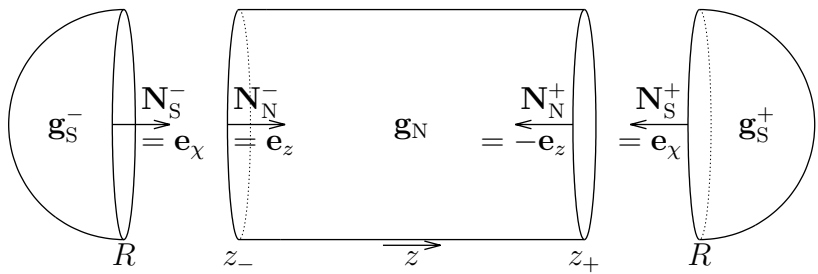}
 \caption{Embedding two stars into Nariai spacetime (schematic)}
 \label{fig:JunctionNariai}
\end{figure}

Now we wish to combine these solutions to a single spacetime
by gluing them along certain boundaries, thereby allowing for
specific discontinuities which are restricted by the 
condition that there shall be no surface layers along the 
identified surfaces in the newly constructed spacetime.
Like in electrodynamics, this results in junction conditions 
the precise form of which were worked out by several people~\cite{Lanczos:1924,Darmois:1927,Israel:1966}, a text-book 
presentation being given in \S\,21.13 of \cite{Misner.Thorne.Wheeler:Gravitation}. In general, these conditions 
state that the induced metrics $\tensor{h}{}$ and extrinsic curvatures 
$\tensor{K}{}$ (essentially corresponding to the normal derivatives of 
the induced metrics) of the hypersurfaces that are to be identified 
have to coincide. In our case, each boundary is the history of an $SO(3)$-orbit, i.e. it is a timelike surface of topology 
$\mathbb{R}\times S^2$. In that special case the junction 
conditions can also be given an alternative form 
\cite{Carrera.Giulini:2010a}, part of which states that the 
areal radii and the Misner-Sharp masses must coincide.  

In charged-Nariai spacetime, at each instant in time,  the 
boundary surfaces are located at $\z$ with $z^+ > 0$ and 
$z^- < 0$ and inward pointing normal 1-form 
$\tensor{N}{N}^\pm=\mp \dd z$. We will embed each star along 
its surface $\chi_b$, defined by the vanishing of the pressure 
$P(\chi_b) = 0$, into charged-Nariai spacetime. Here, the 
outward pointing normal 1-form is 
$\tensor{N}{S}^\pm = \frac{1}{\sqrt{\beta \left( \R{S}^2 \sin^2 \chi + \Ss 
\right)}} \,\dd \chi$. For a schematic representation see 
\fref{fig:JunctionNariai}. For the induced metrics we obtain
\numparts
\begin{eqnarray}
 \tensor{h}{N} &= - \cos^2 \left( \frac{\lambda \z}{\R{N}} \right) \,\dd t^2 
 + \R{N}^2 \,\dd\Omega^2\,,
 \label{eqn:NariaiInduced} \\
 \tensor{h}{S}^\pm &= -\mathcal{A}^\pm \,\frac{F_b^2}{2 \sqrt{11 \beta} F_b + 
 1} \,\dd t^2 + \R{S}^2 \sin^2 \chi_b \,\dd\Omega^2\,, \label{eqn:StarInduced}
\end{eqnarray}
\endnumparts
and for the extrinsic curvatures
\numparts
\begin{eqnarray}
 \fl \tensor{K}{N} &= \pm \frac{\lambda}{\R{N}} \cos \left( 
 \frac{\lambda  \z}{\R{N}} \right) \sin \left( \frac{\lambda \z}{\R{N}} 
 \right) \,\dd t^2\,, \label{eqn:NariaiExtrinsic} \\
 \fl \tensor{K}{S}^\pm &= \sqrt{\beta \left( \R{S}^2 \sin^2 \chi + \Ss 
 \right)} \left( \mathcal{A}^\pm \,\frac{F_b (\sqrt{11 \beta} F_b + 1) 
 F_b'}{(2 \sqrt{11 \beta} F_b + 1)^2} \,\dd  t^2 - \R{S}^2 \sin \chi_b \cos 
 \chi_b \,\dd\Omega^2 \right)\,, \label{eqn:StarExtrinsic}
\end{eqnarray}
\endnumparts
where $F_b = F(\chi_b)$.

Comparing the spatial parts of the extrinsic curvatures we immediately 
get $\chi_b = \frac{\pi}{2}$, so that the star's surface is precisely 
the equator of the 3-sphere. Equality of the spatial parts of the 
induced metric then tells us that the radii of the charged-Nariai 
spacetime $\R{N}$ and the star $\R{S}$ must be the same, 
hence $\R{N}=\R{S}=:\R{}$. Clearly, these results had to be 
expected on geometric grounds.  
These two conditions also ensure that each star has the mass $M=\frac{\R{}}{2G}$, as demanded by the charged-Nariai metric. 
Using the formula \eref{eqn:DerivationF} for the derivative of 
$F$ we can simplify the expression for the extrinsic curvature 
of the star to
\begin{equation}
 \tensor{K}{S}^\pm = \mathcal{A}^\pm \R{} F_b \,\frac{\sqrt{11 \beta} F_b + 
 1}{2 \sqrt{11 \beta} F_b + 1} \,\dd  t^2\,.
\end{equation}
We already know from \sref{sec:ChargedNariai} that the star at 
$z^+$ must have the charge $Q^+(\R{}) = q$ and the other one 
$Q^-(\R{}) = -q$ so that $5\beta\R{}^6=Gq^2$. Since the pressure 
vanishes at the surface, $P(\chi_b) = 0$, we get from 
\eref{eqn:StarPressure} an expression for the cosmological 
constant
\begin{equation} \label{eqn:CosmologicalConstant}
 \Lambda = \alpha - 4 \beta \R{}^2.
\end{equation}
This condition is not independent as it is also a consequence of 
\eref{eqn:NariaiEinsteinA} and the definition of $\R{}$ by 
$1 -\alpha \R{}^2 - \beta \R{}^4 = 0$. However, it allows to 
derive some useful identities from the definition of the 
radius $\R{S}$ and $\alpha$. Indeed, from
\begin{equation}
 2 \sqrt{\alpha^2 + 4 \beta} = 2 \alpha + 4 \beta \R{}^2 = 3 \alpha - 
 \Lambda =  2 \ro
\end{equation}
we obtain $\alpha^2 = \ro^2 - 4 \beta$ and $2 \beta \R{}^2 = \ro - \alpha$. 
Using these relations we can rewrite the pressure function as
\begin{equation} \label{eqn:Pressure}
 \bar{P}(\chi) = \sqrt{\ro - \beta \R{}^2 \cos^2 \chi} \left( \sqrt{11 \beta} 
 + \frac{1}{F(\chi)} \right) \R{} \cos \chi - 2 \beta \R{}^2 \cos^2 \chi\,,
\end{equation}
using $\beta \left( \R{}^2 \sin^2 \chi + \Ss \right) = \ro - \beta \R{}^2 
\cos^2 \chi$. The constant $\mathcal{B}$ is related to the central pressure 
$p_c = P(0)$ by
\begin{equation} \label{eqn:CentralPressure}
  \p = \sqrt{11 \beta} + \frac{1}{F_c} - 2 \beta \R{}^2\,, 
\end{equation}
where $F_c = F(0)$. Solving this for $\mathcal{B}$ we get
\begin{equation} \label{eqn:ConstantB}
 \mathcal{B} = \frac{\p + 2 \beta \R{}^2 + \sqrt{11 \beta}}{\p +  2 \beta 
 \R{}^2 - \sqrt{11 \beta}} \exp \left( -\sqrt{11} \arcsin \left( \sqrt{1 - 
 \frac{4 \beta}{\ro^2}} \right) \right)\,.
\end{equation}
The constant $\mathcal{A}^\pm$ is easily determined by the 
comparison of the time components of the induced metrics as
\begin{equation} 
\label{eqn:ConstantA}
 \mathcal{A}^\pm = \frac{2 \sqrt{11 \beta} F_b +1}{F_b^2} \cos^2 \left( 
 \frac{\lambda \z}{\R{}} \right)\,.
\end{equation}
Finally we compare the time components of the extrinsic curvatures.
This leads to the following relation between the central 
pressure and the position of the star:
\begin{equation} 
\label{eqn:DistancePressureRelation}
\pm \tan \left( \frac{\lambda \z}{\R{}} \right) = \frac{\R{}^2}{\lambda} 
\left( \sqrt{11 \beta} + \frac{1}{F_b} \right)\,.
\end{equation}

Altogether the star is described by three independent parameters similar to 
the Nariai spacetime. We can choose from three independent parameter sets 
describing the charge $(q, \sigma, \beta)$, the mass $(M, \R{}, \ro)$ and 
their positions $\z$ or central pressures $\p$ related by 
\eref{eqn:DistancePressureRelation}.

\subsection{Allowed parameter sets}

In the following we will concentrate on the parameters $\ro$, $\beta$ and 
$\p$ and wish to characterise their allowed domains.

The occurent square root $\sqrt{\ro - \beta \R{}^2 \cos^2 \chi}$ in the 
pressure \eref{eqn:Pressure} is real for all $\chi \le \frac{\pi}{2}$ if the 
mass density is positive, $\ro > 0$, and $\ro - \beta \R{}^2 = 
\frac{1}{\R{}^2} > 0$. The radius $\R{}^2 = \frac{\ro}{2 \beta} \left( 1 - 
\sqrt{1 - \frac{4 \beta}{\ro^2}} \right)$ as a function of $\ro$ and $\beta$ 
is real if $\beta < \frac{\ro^2}{4}$. These equations imply an upper 
bound for the modulus of the charge.

From the stress-energy conservation \eref{eqn:StarConservation} we get for the 
derivative of the pressure the expression
\begin{equation} \label{eqn:PressureDiff}
 \fl \bar{P}'(\chi) = 15 \beta \R{}^2 \sin \chi \cos \chi - \left( \ro 
 + \bar{P}(\chi) \right) \frac{\R{} \sin \chi}{\sqrt{\ro - \beta \R{}^2 \cos^2 
 \chi}} \left( \sqrt{11 \beta} + \frac{1}{F(\chi)} \right)\,, 
\end{equation}
so that the derivative at the boundary is negative
\begin{equation}
 \bar{P}_b' = - \R{} \sqrt{\ro} \left( \sqrt{11 \beta} + \frac{1}{F_b} \right) 
 = \mp \sqrt{\ro} \,\frac{\lambda}{\R{}} \tan \left( \frac{\lambda \z}{\R{}} 
 \right) < 0\,.
\end{equation}
This implies that just below the star's surface the pressure is 
positive. But since the star's surface was defined to be the first zero 
of the pressure, the pressure must be positive everywhere within the 
star. In fact, we can show that the pressure is positive if and only 
if  the central pressure is positive. In the neutral 
case, $\beta = 0$, this is immediate since only the negative term in \eref{eqn:PressureDiff} remains so that the pressure must be 
positive and monotonically decreasing. In the charged case this is 
a little harder to see. From \eref{eqn:Pressure} we obtain the inequality
\begin{equation}
 \frac{2 \beta \R{} \cos \chi}{\sqrt{\ro - \beta \R{}^2 \cos^2 \chi}} \le 
 \sqrt{11 \beta} + \frac{1}{F(\chi)}\,.
\end{equation}
We first notice that the right-hand side is monotonically increasing 
in $\p$ because of $\frac{\mathrm{d}}{\mathrm{d} \p} \frac{1}{F(\chi)} > 0$, as one easily verifies by direct calculation. Therefore, we may 
set $\p = 0$, in which case $F(\chi)$ is monotonically decreasing 
in $\chi$. Considering \eref{eqn:DerivationF} for the derivative 
of $F(\chi)$ this means that $2\sqrt{11 \beta}F_c+1=\frac{2 \beta \R{}^2 + \sqrt{11 \beta}}{2 \beta \R{}^2 - \sqrt{11 \beta}} < 0$ or $2 \beta \R{}^2 < \sqrt{11 \beta}$ using \eref{eqn:CentralPressure} with $\p = 0$. 
The last inequality can be rewritten as $1 - \sqrt{1-4x} < \sqrt{11x}$ 
with $x = \frac{\beta}{\ro^2}$ which is true for $0 < x \le \frac{1}{4}$. 
Hence we have
\begin{equation}
 \sqrt{11 \beta} + \frac{1}{F(\chi)} \ge \sqrt{11 \beta} + \frac{1}{F_c} = 2 
 \beta \R{}^2.
\end{equation}
Since the left side is monotonically decreasing in $\chi$, which 
can again be checked easily by direct calculation, we have
\begin{equation}
 \frac{2 \beta \R{} \cos \chi}{\sqrt{\ro - \beta \R{}^2 \cos^2 \chi}} \le 2   
 \beta \R{}^2\,.
\end{equation}
In total we thus get 
\begin{equation}
 \frac{2 \beta \R{} \cos \chi}{\sqrt{\ro - \beta \R{}^2 \cos^2 \chi}} \le 2 
 \beta \R{}^2 \le \sqrt{11 \beta} + \frac{1}{F(\chi)}\,,
\end{equation}
showing the desired result that the pressure is positive everywhere 
within in the star. 
Note that it is not excluded that the pressure increases near 
the centre in an outward direction because there we have
\begin{equation}
 \bar{P}(\chi) = \p + \frac{1}{2} \left( 15 \beta \R{}^2 - \R{} \frac{\left( 
 \ro + \p \right)}{\sqrt{\ro}} \left( \p + 2 \beta \R{}^2 \right) \right) 
 \chi^2 + \mathcal{O}(\chi^3).
\end{equation}

\begin{figure}[b]
 \centering
 \resizebox{0.58\columnwidth}{!}{\input{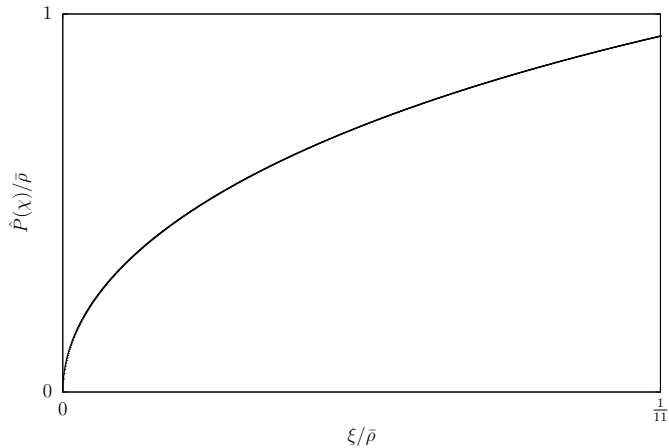}}
 \caption{Maximal pressure}
 \label{fig:MaxPressure}
\end{figure}

Having shown that the pressure is bounded below by zero we next 
wish to show that it is also bounded above by the density. 
As already stated, the pressure may assume its maximal value 
off the centre. If the maximum is at the centre, the pressure 
must monotonically decrease as we move away from the centre 
towards the surface. Hence $\bar{P}(\chi) \leq \p$. Assuming 
there exists a maximum at $\hat{\chi} \in (0,\pi/2)$ bigger than 
the central pressure, we have 
$0$ 
\begin{equation}
 \fl \bar{P}'(\hat{\chi}) = 15 \beta \R{}^2 \sin \hat{\chi} \cos \hat{\chi} - 
 \left( \ro + \bar{P}(\hat{\chi}) \right) \frac{\R{} \sin \hat{\chi}}{\sqrt{\ro 
 - \beta \R{}^2 \cos^2 \hat{\chi}}} \left( \sqrt{11 \beta} + 
 \frac{1}{F(\hat{\chi})} \right) = 0.
\end{equation}
If we insert  
\begin{equation}
 \fl \hat{P} := \bar{P}(\hat{\chi}) = \sqrt{\ro - \beta \R{}^2 \cos^2 
 \hat{\chi}} \left( \sqrt{11 \beta} + \frac{1}{F(\hat{\chi})} \right) \R{} \cos 
 \hat{\chi} - 2 \beta \R{}^2 \cos^2 \hat{\chi}
\end{equation}
we obtain
\begin{equation}
 15 \beta \R{}^2 \cos^2 \hat{\chi} \left( \ro - \beta \R{}^2 \cos^2 \hat{\chi} 
 \right) = \left( \ro + \hat{P} \right) \left( \hat{P} + 2 \beta \R{}^2 \cos^2 
 \hat{\chi} \right)\,.
\end{equation}
Using the abbreviation $\xi = \beta \R{}^2 \cos^2 \hat{\chi}$ we can 
rewrite this equation as
\begin{equation}
 \hat{P}^2 + \left( \ro + 2\xi \right) \hat{P} + \left( 15\xi - 13 \ro \right) 
 \xi = 0
\end{equation}
with the positive solution
\begin{equation}
 \hat{P} = - \left(\frac{\ro}{2} + \xi \right) + \sqrt{ \left(\frac{\ro}{2} + 
 \xi 
 \right)^2 + \left( 13 \ro - 15\xi \right) \xi}.
\end{equation}
As we will see below, the relevant sector is given by 
$0 \leq \beta \leq \frac{10}{121} \ro^2$. Hence we have $0 \leq 
\xi \leq \frac{1}{11} \ro$, leading to the desired bound 
$\bar{P} \leq \ro$. This is also shown in \fref{fig:MaxPressure}.

Since the derivative is bounded from above the pressure never reaches 
infinity before it decreases. Moreover, if the central pressure is 
negative there must always be at least one sphere within the star 
where the pressure either vanishes (contradicting the assumption that 
the radius is the first zero) or diverges, leading to a discontinuity.

\begin{figure}[b]
\centering
\input{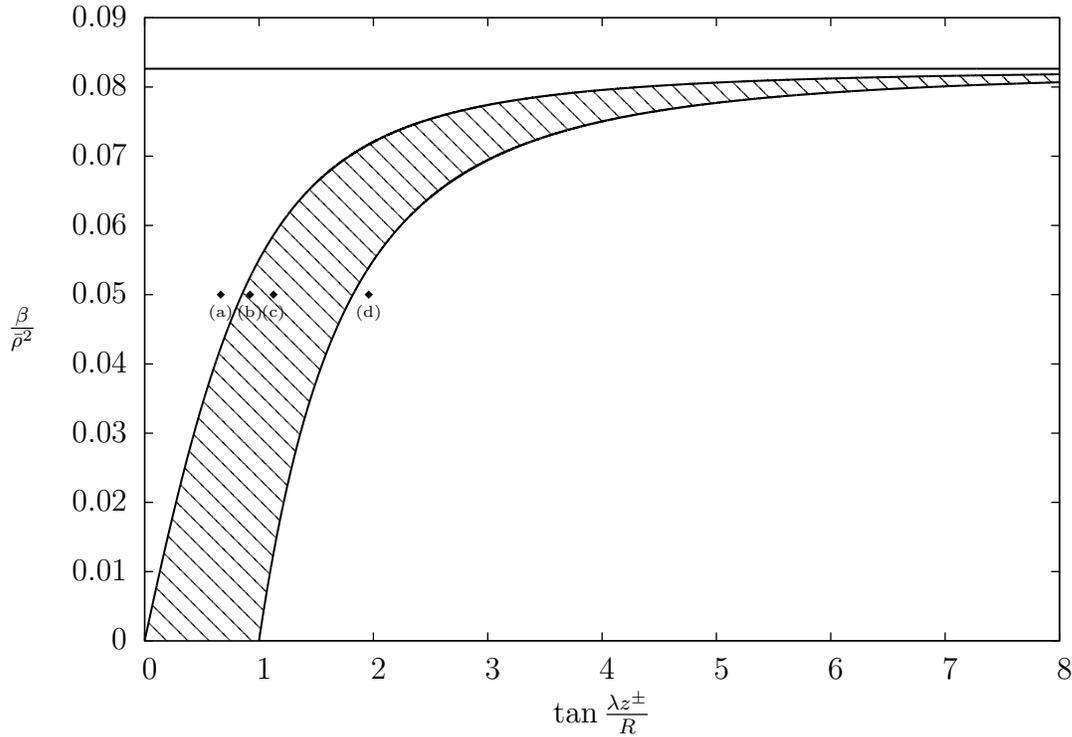}
\caption{In the ruled region the pressure is positive within the star. The rule 
corresponds to the critical charge $\beta/\ro^2=\frac{10}{121}$. The four marks 
correspond to the four pressure distributions shown in 
\fref{fig:PressureDistribution}.}
\label{fig:charge-distance}
\end{figure}

\begin{figure}[b]
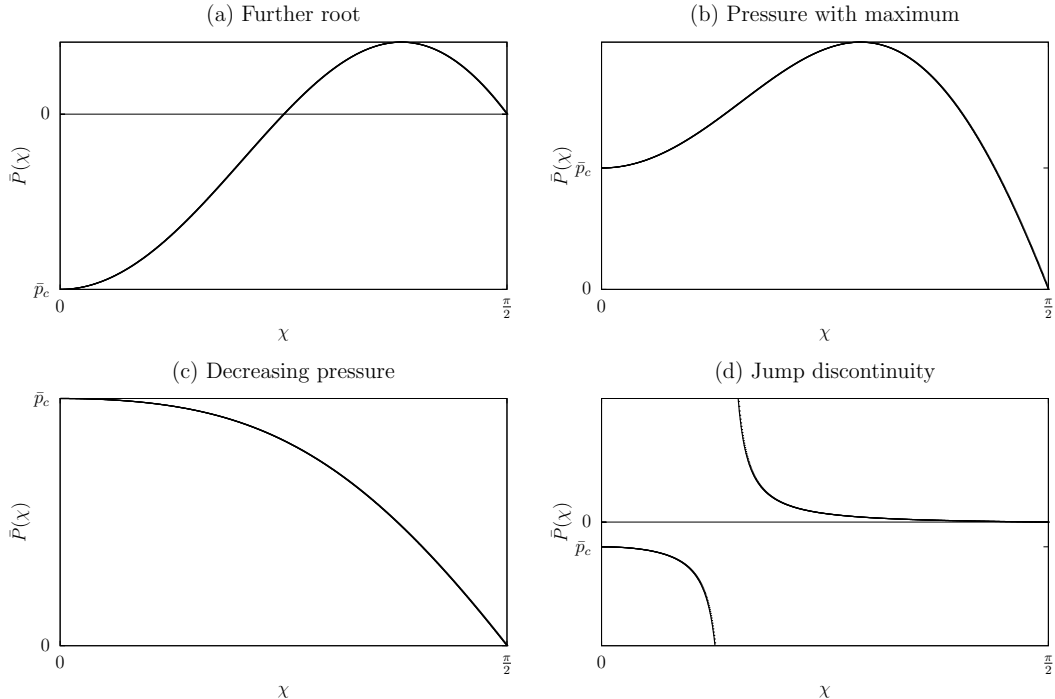

 \centering
 \begin{minipage}[b]{0.45\textwidth}
%  \caption{Further root}
  \resizebox{\columnwidth}{!}{\input{figure4a.tex}}
 \end{minipage}
 \begin{minipage}[b]{0.45\textwidth}
%  \caption{Pressure with maximum}
  \resizebox{\columnwidth}{!}{\input{figure4b.tex}}
  \end{minipage} \\
 \begin{minipage}[b]{0.45\textwidth}
%  \caption{Decreasing pressure}
  \resizebox{\columnwidth}{!}{\input{figure4c.tex}}
 \end{minipage}
 \begin{minipage}[b]{0.45\textwidth}
%  \caption{Jump discontinuity}
  \resizebox{\columnwidth}{!}{\input{figure4d.tex}}
 \end{minipage}
 \caption{Characteristic pressure distributions (their positions are marked 
 in \fref{fig:charge-distance})}
 \label{fig:PressureDistribution}
\end{figure}

From the previous section we know that the central pressure is related to the 
position of the star. So not all positions are allowed, as shown in 
\fref{fig:charge-distance}. For each position there is only an interval of 
possible charges of the star so that the pressure is positive everywhere. In 
the other regions the central pressure is negative and discontinuities 
(lower right region) or further roots of the pressure function (upper left 
region) occur. Some characteristic pressure distributions are shown in 
\fref{fig:PressureDistribution}.

As \fref{fig:charge-distance} indicates there is a critical upper 
bound for the modulus of the charge, given by 
$\beta_\mathrm{crit} < \frac{1}{4} \ro^2$. This is the point where 
$\lambda = 0$ or $\beta = \frac{10}{121} \ro^2$ and the charged-Nariai spacetime changes its topology, turning into a cosmological Bertotti-Robinson spacetime. If we keep the central pressure $\p$ constant and increase the charge we move within the allowed region but $\mathrm{g}_{00}$ tends to 
zero. The latter implies diverging acceleration of the stationary 
Killing orbits and hence an instability of the star. Higher charges are
also not possible, for the pressure then turns negative. Because of the junction condition
\begin{equation}
 \mp \tanh \left( \frac{\lambda \z}{\R{}} \right) = \frac{\R{}^2}{\lambda}    
 \left( \sqrt{11 \beta} + \frac{1}{F_b} \right)
\end{equation}
we have $\sqrt{11 \beta} + \frac{1}{F_b} < 0$ and thus $\bar{P}(\chi) < 0$. 
However, it should be possible to embed only one star.

%Neutral limit
%%%%%%%%%%%%%%%%%%%%%%%%%%%%%%%%%%%%%%%%%%%%%%%%%%%%%%%%%%%%
\section{Neutral Limit} \label{sec:NeutralLimit}

Now we want to consider the neutral limit $\beta \rightarrow 0$. 
For this we have to keep two parameters constant. These will be 
the mass density $\ro$ and the central pressure $\p$. 

At first we consider the charged-Nariai spacetime and start with the radius 
$\R{}$. Because of
\begin{equation}
 2 \beta \R{}^2 = \ro \left( 1- \sqrt{1 - \frac{4 \beta}{\ro^2}} \right) = 
 \frac{2 \beta}{\ro} + \frac{2 \beta^2}{\ro^3} + \0{3}
\end{equation}
we obtain $\R{}^2= \frac{1}{\ro} + \0{}$. We can derive the Taylor expansion 
for the cosmological constant from \eref{eqn:CosmologicalConstant}. This gives 
us 
\begin{equation}
 \Lambda = \ro \left( 3 \sqrt{1 - \frac{4 \beta}{\ro^2}} - 2 \right) = \ro - 
 \frac{6 \beta}{\ro} + \0{2}
\end{equation}
and agrees with the expression we would get from \eref{eqn:NariaiEinsteinA}. 
Furthermore we have
\begin{equation}
 \lambda^2 = 1 - 10 \beta \R{}^4 = 1 - \frac{10 \beta}{\ro^2} + \0{2}
\end{equation}
and thus $\lambda = 1 + \0{}$. Hence the charged Nariai metric reduces to the 
common Nariai metric~\cite{Nariai:1999-a,Nariai:1999-b}
\numparts
\begin{eqnarray}
 \tensor{g}{N} &= - \cos^2 \left( \frac{z}{{\R{}}_0} \right) \,\dd t^2 + \dd 
 z^2 + {\R{}}_0^2 \,\dd\Omega^2 \\
 &= \frac{1}{\Lambda_0} \left( -\cos^2(Z) \,\dd T^2 + \dd Z^2 + \dd\Omega^2 
 \right)
\end{eqnarray}
\endnumparts
with $T = \sqrt{\Lambda_0} \,t$, $Z = \sqrt{\Lambda_0} \,z$ and $\Lambda_0 = 
1/{\R{}}_0^2 = \ro$.

Now lets turn to the star's metric. We begin with the expansion of the 
constant $\mathcal{B}$. We have for the first part in \eref{eqn:ConstantB}
\begin{equation}
 \frac{\p + 2 \beta \R{}^2 + \sqrt{11 \beta}}{\p + 2 \beta \R{}^2 - \sqrt{11 
 \beta}} = 1 + \frac{2}{\p} \sqrt{11 \beta} + \0{}\,.
\end{equation}
For the second part we consider (note: $\arcsin(1-x) = \frac{\pi}{2} - 
\sqrt{2x} + \mathcal{O}\left( x^{3/2} \right)$)
\begin{equation}
 \sqrt{11} \arcsin \left( 1 - \frac{2 \beta \R{}^2}{\ro} \cos^2 \chi \right) = 
 \sqrt{11} \,\frac{\pi}{2} - \frac{2}{\ro} \,\sqrt{11 \beta} \cos \chi + \0{}\,,
\end{equation}
so that 
\begin{equation} \fl
 \exp \left[\sqrt{11} \arcsin \left( 1 - \frac{2 \beta \R{}^2}{\ro} \cos^2 
 \chi \right) \right] = \exp \left( \sqrt{11} \,\frac{\pi}{2} \right) \left[ 1 
 + \frac{2}{\ro} \,\sqrt{11 \beta} \cos \chi + \0{} \right].
\end{equation}
If we set $\chi = \frac{\pi}{2}$ and invert the expression we can derive the 
expansion for $\mathcal{B}$ 
\begin{equation}
 \mathcal{B} = \exp \left( -\sqrt{11} \,\frac{\pi}{2} \right) \left( 1 + 
 \frac{2}{\ro} \,\sqrt{11 \beta} \,\frac{\ro + \p}{\p} + \0{} \right)\,.
\end{equation}
Combining the last two results we can expand the function $F(\chi)$
\begin{eqnarray} 
 2 \sqrt{11 \beta} F(\chi) = \frac{2}{\ro} \sqrt{11\beta} \left( \frac{\ro + 
 \p}{\p} - \cos \chi \right) + \0{}\,,
\end{eqnarray}
so that
\begin{equation}
 \frac{F^2(\chi)}{2 \sqrt{11 \beta} F(\chi) + 1} = \frac{1}{\ro^2} \left[
 \frac{\ro + \p}{\p} - \cos \chi \right]^2 +  \0{1/2}\,.
\end{equation}
Before we can determine the metric we need the expansion of $\mathcal{A}$ and 
$\z$. We get
\begin{eqnarray}
 \z &= \pm \frac{\R{}}{\lambda} \arctan \left[\frac{\R{}^2}{\lambda} \left( 
 \sqrt{11 \beta} + \frac{1}{F_b} \right) \right] \nonumber \\
 &= \pm \left( \frac{1}{\sqrt{\ro}} + \0{1/2} \right) \arctan \left[ 
 \frac{\p}{\ro + \p} + \0{1/2} \right] \\
 &= \pm \frac{1}{\sqrt{\ro}} \arctan \left( \frac{\p}{\ro + \p} \right) + 
 \0{1/2}\,. \nonumber
\end{eqnarray}
Hence $\z_0 = \pm {\R{}}_0 \arctan \frac{\p}{\ro + \p}$. Furthermore we have 
$\cos^2 \left( \frac{\lambda \z}{\R{}} \right) = \cos^2 \left( 
\frac{\z_0}{{\R{}}_0} \right) + \0{1/2}$. Now we can easily calculate the 
constant $\mathcal{A}$ from \eref{eqn:ConstantA}
\begin{equation}
 \mathcal{A}^\pm = \ro^2 \left[ \frac{\p}{\ro + \p} \right]^2 \cos^2 \left( 
 \frac{\z_0}{{\R{}}_0} \right) + \0{1/2}.
\end{equation}
Combining all our results, we can expand the metric components
\numparts
\begin{eqnarray}
 \mathrm{g}_{00} &= \cos^2 \left( \frac{\z_0}{{\R{}}_0} \right) \left[ 1 - 
 \frac{\p}{\ro + \p} \cos \chi \right]^2 + \0{1/2}\,, \\
 \mathrm{g}_{11} &= \frac{1}{\ro - \beta \R{}^2 \cos^2 \chi} = \frac{1}{\ro} + 
 \0{}\,.
\end{eqnarray}
\endnumparts
Hence the star's metric in neutral limit is
\begin{equation}
 \tensor{g}{S} = -\cos^2 \left( \frac{\z_0}{{\R{}}_0} \right) \left[ 1 - 
 \frac{\p}{\ro + \p} \cos \chi \right]^2 \dd t^2 + \frac{1}{\ro} \left( \dd 
 \chi^2 + \dd \Omega^2 \right)\,.
\end{equation}
This is an agreement with eq.\,(3.33) of~\cite{Boehmer:2004},
except for the prefactor which depends on the surrounding spacetime. 
Upon rescaling the time coordinate it is possible to arrive at 
the same expression.

Finally we consider the pressure
\begin{equation}
 \bar{P}(\chi) 
=\ro\,\frac{\p \cos \chi}{\ro + \p - \p \cos \chi} + \0{1/2}\,,
\end{equation}
which for $\beta=0$ coincides with eq.\,(3.32) of \cite{Boehmer:2004}.

%%%%%%%%%%%%%%%%%%%%%%%%%%%%%%%%%%%%%%%%%%%%%%%%%%%%%%%%%%%%
\section{Discussion}
In this paper we investigated exact solutions to Einstein's 
equations which represent two spherically symmetric stars 
made from an incompressible perfect fluid, possibly with 
non-vanishing charge density which is constant with respect 
to the areal volume element. The stars are kept at constant 
distance and the solution is globally static and spherically 
symmetric. The topology of 
the spatial splices of simultaneity is that of a three-sphere, 
so that the total electric charge must be zero. In fact, the 
stars have charges of equal modulus and opposite sign, leading 
to further attraction. The combined gravitational and electric 
attraction of the stars is balanced by the negative pressure 
of a positive cosmological constant without causing a 
cosmological horizon separating the stars. These solutions 
were not contained in previous analyses and somehow bridge 
between the results obtained in \cite{UzanEllisLarena:2011}
and those in \cite{Boehmer:2004}. Being exact solutions to 
Einstein's equations it is clear that they can claim some interest 
in their own right. In our interpretation of the possible physical significance of these solutions we follow 
\cite{UzanEllisLarena:2011}, who see the study of exact 
two-mass solutions as an initial step towards a more 
rigorous understanding of local geometric structure in 
inhomogeneous cosmologies, though this admittedly means stressing 
one's imagination, in particular given the ratios of the 
cosmological constant to the mass density involved in our 
solutions. However, whereas it seems now clear that 
the spin-spin-interaction of aligned (sub-extremal) Kerr 
black-holes cannot balance their gravitational attraction 
(so as to result in a stationary exterior spacetime)~\cite{Neugebauer:Henning:2009}, 
it is interesting to see what a cosmological constant can do. 
This, to us, motivates further investigations in this direction
and perhaps combine results. 
\ack
Hospitality and support of the Center of Applied Space Technology and
Microgravity (ZARM) at Bremen is gratefully acknowledged.  
Micheal Fennen was supported by a Ph.D. grant of the German Research 
Foundation (DFG) within its Research Training  Group no.\,1620 
\emph{Models of Gravity}. Domenico Giulini was supported by the 
Cluster of Excellence \emph{Centre for Quantum Engineering and Space-Time Research} 
of the DFG.

\appendix
\setcounter{section}{1}
\section{Curvatures of Nariai-Type metrics}
We consider metrics of the form \eref{eq:SpehricalSymmMetric-2}
and wish to calculate the curvature coefficients with respect to 
the orthonormal tetrad \eref{eq:OrthonormalFrame-0}-\eref{eq:OrthonormalFrame-3}.
This we do by solving Cartan's first structure equation
(expressing vanishing torsion)
\begin{equation}
\label{eq:CartanFirst}
\dd\theta^a+\omega^a_{\phantom{a}b}\wedge\theta^b=0\,,
\end{equation} 
for the connection 1-forms $\connection{a}{b}$.
This gives a unique solution for metric-compatible 
connections, which satisfy 
$g_{ac}\connection{c}{b}=-g_{bc}\connection{c}{a}$.
Note that here all indices refer to components with 
respect to the orthonormal tetrad, so that 
$g_{11}=g_{22}=g_{33}=-g_{00}=1$ and $g_{ab}=0$ for $a\ne b$.
Using \eref{eq:OrthonormalFrame-0}-\eref{eq:OrthonormalFrame-3}
a straightforward calculation gives
\begin{eqnarray}
\label{eq:ConnectionCoeff-01}
\connection{0}{1}&=&\frac{a'}{a}\theta^0=a'\,\dd t\,,\\
\label{eq:ConnectionCoeff-02}
\connection{0}{2}&=&\connection{0}{3}= 0\,,\\
\label{eq:ConnectionCoeff-12}
\connection{1}{2}&=& -\frac{R'}{R}\ \theta^2=-R'\,\dd \vartheta\,,\\
\label{eq:ConnectionCoeff-13}
\connection{1}{3}&=& -\frac{R'}{R}\ \theta^3=-R'\,\sin\vartheta\,\dd\varphi\\
\label{eq:ConnectionCoeff-23}
\connection{2}{3}&=& -\frac{\cot\vartheta}{R}\ \theta^3=-\cos\vartheta\,\dd \varphi\,.
\end{eqnarray}
From these the curvature 2-forms follow by Cartan's second structure
equation: 
\begin{equation}
\label{eq:CartanSecond}
\curvature{a}{b}
=\frac{1}{2}\,R^a_{\phantom{a}b\,cd}\ \theta^c\wedge\theta^d
=\dd \connection{a}{b}+\connection{a}{c}\wedge\connection{c}{b}
\end{equation} 
and thus, in turn, the non-vanishing components of the 
(totally covariant) Riemann tensor 
$R_{ab\,cd}=g_{an}R^n_{\phantom{n}b\,cd}$:
\begin{eqnarray}
\label{eq:RiemanntensorComp-01}
R_{01\,01}&=&\frac{a''}{a}\,,\\
\label{eq:RiemanntensorComp-02}
R_{02\,02}&=&R_{03\,03}=\frac{a'R'}{aR}\,,\\
\label{eq:RiemanntensorComp-12}
R_{12\,12}&=&R_{13\,13}=-\frac{R''}{R}\,,\\
\label{eq:RiemanntensorComp-23}
R_{23\,23}&=&\frac{1-R'^2}{R^2}\,.
\end{eqnarray} 

The components of the Ricci tensor follow from 
\begin{equation}
\label{eq:Riccitensor}
R_{ab}=-R_{0a\,0b}+R_{1a\,1b}+R_{2a\,2b}+R_{3a\,3b}\,,
\end{equation} 
of which the non-vanishing ones are 
\begin{eqnarray}
\label{eq:RiccitensorComp-00}
R_{00}&=&\frac{a''}{a}+2\,\frac{a'R'}{aR}\,,\\
\label{eq:RiccitensorComp-11}
R_{11}&=&-\frac{a''}{a}-2\,\frac{R''}{R}\,,\\
\label{eq:RiccitensorComp-22}
R_{22}&=&R_{33}=\frac{1-R'^2}{R^2}-\frac{a'R'}{aR}-\frac{R''}{R}\,.
\end{eqnarray} 
This gives the scalar curvature
\begin{eqnarray}
\label{eq:ScalarCurvature}
R
&=&-R_{00}+R_{11}+R_{22}+R_{33}\nonumber\\
\label{eq:ScalarCurvature}
&=&2\left[
\frac{1-R'^2}{R^2}-\frac{a''}{a}-2\,\frac{R''}{R}-2\,-\frac{a'R'}{aR}
\right]\,.
\end{eqnarray}

Finally, \eref{eq:RiccitensorComp-00}-\eref{eq:RiccitensorComp-22} and \eref{eq:ScalarCurvature} give the non-vanishing components 
of the Einstein tensor:

\begin{eqnarray}
\label{eq:EinsteinTensorComp-00}
G_{00}&=&R_{00}+\frac{R}{2}=-2\,\frac{R''}{R}+\frac{1-R'^2}{R^2}\,,\\
\label{eq:EinsteinTensorComp-11}
G_{11}&=&R_{11}-\frac{R}{2}=2\,\frac{a'R'}{aR}-\frac{1-R'^2}{R^2}\,,\\
\label{eq:EinsteinTensorComp-22}
G_{22}&=&G_{33}=R_{22}-\frac{R}{2}=\frac{a''}{a}+\frac{R''}{R}+\frac{a'R'}{aR}\,.
\end{eqnarray} 

\newpage
\bibliographystyle{plain}
\bibliography{RELATIVITY,HIST-PHIL-SCI,MATH,QM}

\end{document}